\documentclass[11pt,twoside]{article}
\usepackage{asp2014}

\newcommand{\Ms}{M$_{\odot}$}
\newcommand{\kepler}{\textit{Kepler}}
\aspSuppressVolSlug
\resetcounters

\begin{document}

\title{An overview of the properties of a sample of newly-identified magnetic chemically peculiar stars in the \textit{Kepler} field}
\author{Z. Mikul\'{a}\v{s}ek,$^1$ E. Paunzen,$^1$ S. H\"{u}mmerich,$^{2,3}$ J. Jan\'{i}k,$^1$ K. Bernhard,$^{2,3}$ J.~Krti\v{c}ka,$^1$ and I.~A.~Yakunin,$^4$}
\affil{$^1$Department of Theoretical Physics and Astrophysics, Masaryk University, Kotl\'a\v{r}sk\'a 2, CZ 611\,37 Brno, Czech Republic; \email{mikulas@physics.muni.cz}}
\affil{$^2$Bundesdeutsche Arbeitsgemeinschaft f{\"u}r Ver{\"a}nderliche Sterne e.V. (BAV), D-12169 Berlin, Germany}
\affil{$^3$American Association of Variable Star Observers (AAVSO), 49 Bay State Rd, Cambridge, MA 02138, USA}
\affil{$^4$Special Astrophysical Observatory of the RAS, Nizhnii Arkhyz, Karachai-Cherkessian Republic, 369167, Russia}

\paperauthor{Zden\v{e}k Mikul\'{a}\v{s}ek}{mikulas@physics.muni.cz}{}{Masaryk University}{Department of Theoretical Physics and Astrophysics}{Brno}{}{CZ-61137}{Czech Republic}
\paperauthor{Ernst Paunzen}{epaunzen@physics.muni.cz}{0000-0002-3304-5200}{Masaryk University}{Department of Theoretical Physics and Astrophysics}{Brno}{}{CZ-61137}{Czech Republic}
\paperauthor{Stefan H\"{u}mmerich}{ernham@rz-online.de}{}{American Association of Variable Star Observers}{}{Cambridge}{}{}{USA}
\paperauthor{Jan Jan\'{i}k}{honza@physics.muni.cz}{0000-0002-6384-0184}{Masaryk University}{Department of Theoretical Physics and Astrophysics}{Brno}{}{CZ-61137}{Czech Republic}
\paperauthor{Klaus Bernhard}{klaus.bernhard@liwest.at}{0000-0002-0568-0020}{American Association of Variable Star Observers}{}{Cambridge}{}{}{USA}
\paperauthor{Ji\v{r}\'{i} Krti\v{c}ka}{krticka@physics.muni.cz}{0000-0001-6322-0236}{Masaryk University}{Department of Theoretical Physics and Astrophysics}{Brno}{}{CZ-61137}{Czech Republic}
\paperauthor{Yakunin Ilya}{elias@sao.ru}{0000-0001-7050-7294}{Russian Academy of Science}{Special Astrophysical Observatory}{}{Nizhny Arkhyz}{RU-369167}{Russia}

\begin{abstract}
We present a comprehensive overview of the properties of a sample of 41 magnetic chemically peculiar stars that have been recently identified in the \textit{Kepler}\ field by our team (H\"{u}mmerich et al. 2018). The stars populate the whole age range from zero-age to terminal-age main sequence in the mass interval from 1.5 to 4 M$_{\odot}$. Several of the studied objects exhibit a hitherto unobserved wealth of detail in their light curves indicative of persisting complex surface structures. Monoperiodic variability and light curve stability were identified as cardinal criteria for selecting mCP star candidates among early-type objects in photometric surveys. Subsequent studies will be concerned with an exhaustive follow-up analysis of the new mCP stars, which we expect to lead to new insights on the physics of the CP star phenomenon.
\end{abstract}
\section{Introduction}

Chemically peculiar (CP) stars of the upper main sequence are characterized by specific anomalies in the photospheric abundances of some chemical elements, which are often unevenly distributed on the stellar surface. The group of the magnetic chemically peculiar (mCP) stars consists of the classical Bp/Ap/Fp (CP2) stars and the He-weak/strong (CP4--7) stars \citep{preston,maitzen}. These objects exhibit strictly periodic light, spectral, and magnetic variations that can be adequately explained by the model of a rigidly rotating star with persistent surface structures and a stable global magnetic field that stabilizes the outer parts of the star \citep{Deutsch70}.

The abundance anomalies result from the interplay between radiative levitation and gravitational settling. The observed light variations are caused by the redistribution of flux in the abundance patches (line and continuum blanketing; for details see in \citet{Krti12}, and references therein). Photometric spots are usually bright in the optical region and dark in the far ultraviolet (UV), where the overabundant elements (mostly Si, Fe, rare-earth elements) have plenty of \textit{b-b} and \textit{b-f} transitions. According to convention, photometrically variable mCP stars are referred to, after their bright prototype, as $\alpha^{2}$ Canum Venaticorum (ACV) variables.

Up to the present, the majority of CP stars has been identified as such through an investigation of their spectra \citep[or, rarely, by using $\Delta$$a$ photometry,
][]{Paunz05}, including the determination of the effective temperature, the type of peculiarity and the subtypes according to the overabundance/underabundance of some elements with conspicuous spectral lines. Spectropolarimetry has been used to check for the presence of a magnetic field \citep{Bych09}.

Photometric observations were conducted typically after the spectroscopic detection, mostly to determine or improve the rotational period. Although the observed amplitudes of the light variations in mCP stars are small and do not exceed 0.12\,mag in $V$, they are more readily suited to derive this parameter than spectroscopic observations. It has been shown that mCP star light curves are smooth and characterized by a single or double wave \citep{mathys,mikzoo,Dukes18} that can be satisfactorily well approximated by second-order harmonic polynomials corresponding to a rotating model with one or two dominant photometric spots. As a rule, light curve shapes and periods persist for decades in these stars \citep{Zizn94}.

Recently, photometric observations have become accurate enough to search for new mCP star candidates by inspection of light curve morphology. To this end, highly precise photometric measurements are needed, which are currently provided from space by specialized observatories such as \kepler\ or CoRoT that continuously observe selected areas in the sky. We have carried out a search for new mCP stars in the \kepler\ field to investigate their photometric variability properties using unprecedentedly precise light curves.

\articlefigure[width=0.85\textwidth]{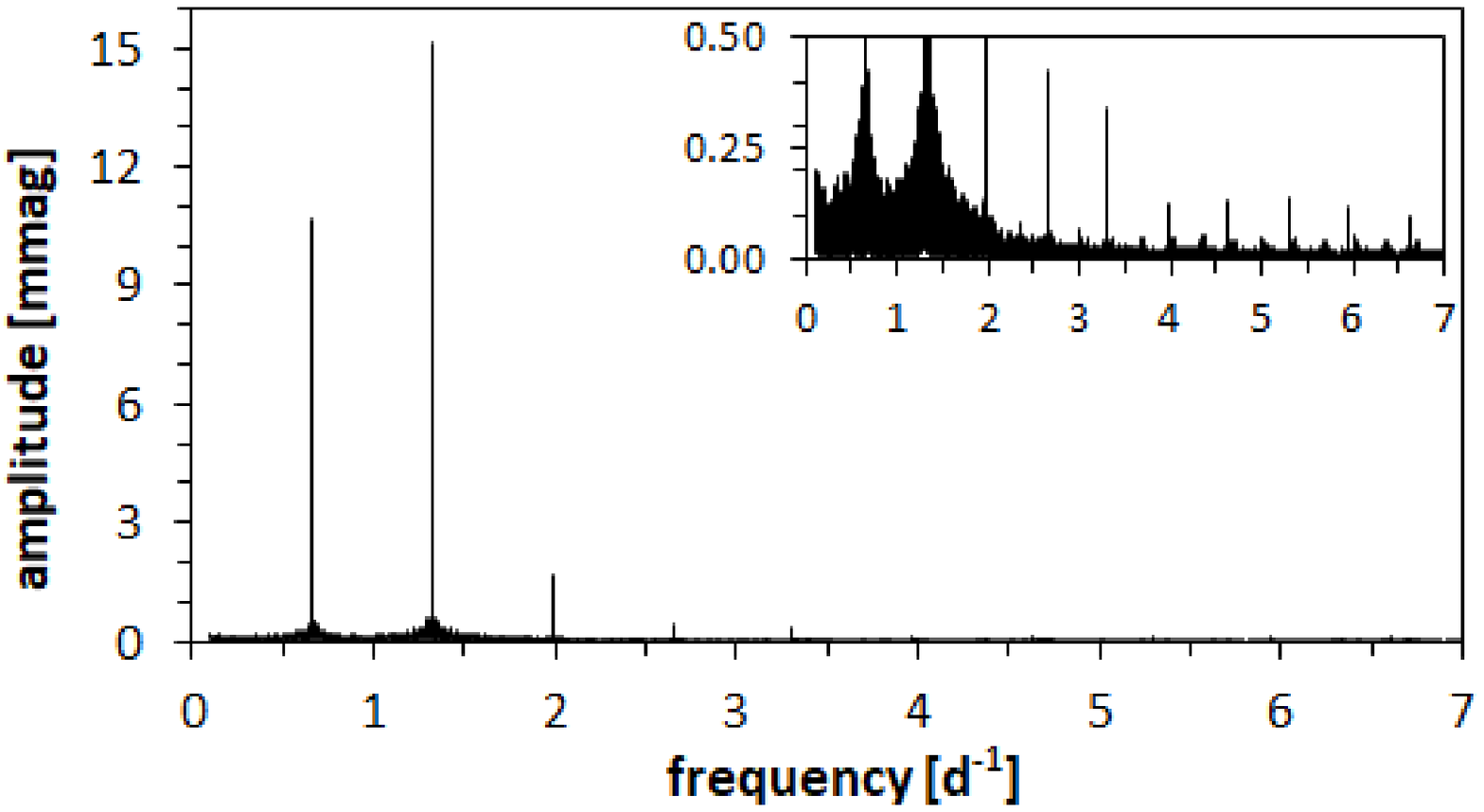}{spektrum}{Fourier amplitude spectrum of the CP2 star KIC\,6950556 (spectrum A0V\,Si), based on non-detrended \kepler\ PDC flux. The spectrum is indicative of a single period $P$\,=\,1.511\,785\,05(4)~d and its harmonics. The inset shows a zoomed-in version with the three main peaks off scale, highlighting the corresponding harmonics; axes are the same as in the main plot.}

\section{Search for mCP stars in the \kepler\ field}\label{criteria}

Unfortunately, in the part of the sky observed during the \kepler\ primary mission, only few mCP stars are known, which has been our motivation to search for additional candidates by non-spectroscopic means. Some compulsory criteria for identifying mCP star candidates were formulated, which are:

\begin{itemize}
	\item spectral type between late B and early F, corresponding color index or effective temperature,
	\item period longer than 0.5 d,
	\item frequency spectrum indicative of a single independent variability frequency with corresponding harmonics,
	\item stable or marginally changing shape of the light curve (see Fig.\,\ref{spektrum}),
	\item light variability amplitude of several hundredths of a magnitude, or less.
\end{itemize}

\articlefiguretwo{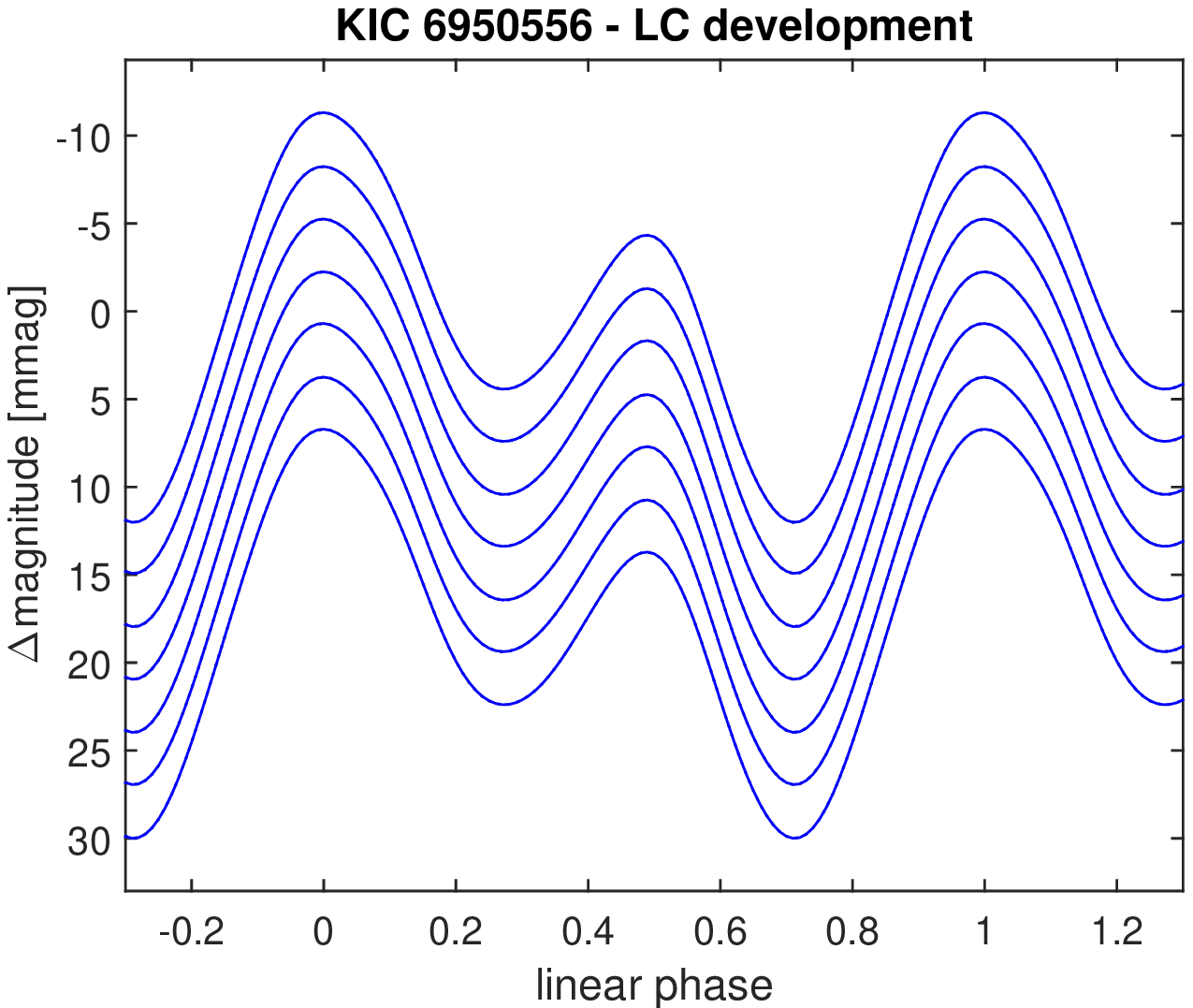}{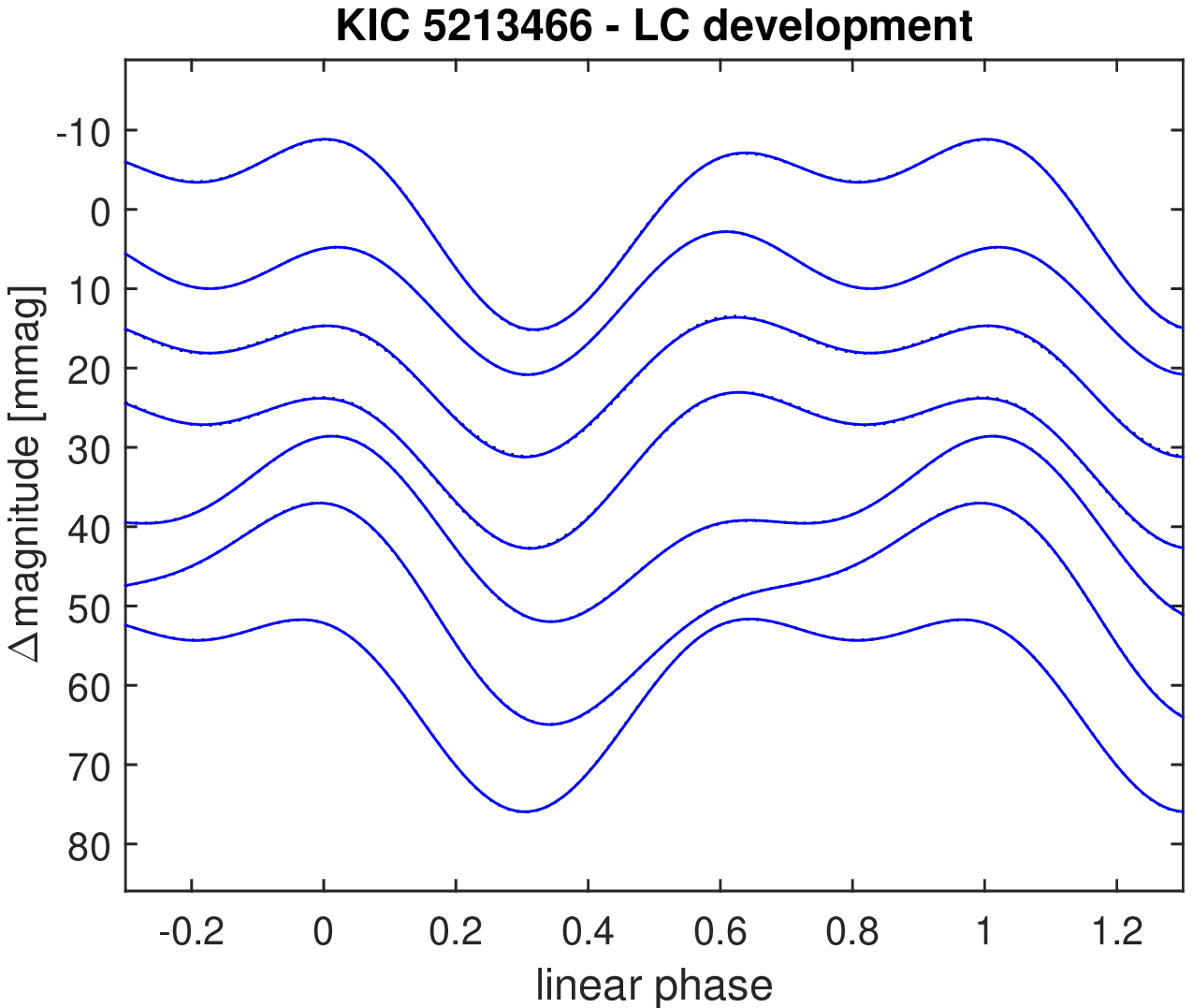}{develop}{Light curve development of the CP2 star KIC\,6950556 (A0V\,Si; left panel) and the chemically normal star KIC\,5213466 (A1V; right panel). The shown light curves are separated by the time-distance of half a year. While the CP2 star does not exhibit any apparent light curve changes throughout the covered time span, the light curve shape of the chemically normal A star displays apparent variations of unknown nature.}

Crucial for confirming a new mCP star candidate is, of course, the spectroscopic identification and determination of spectral type and peculiarity subtype. To this end, newly acquired and archival spectra were employed, which enabled us to confirm most of our candidates as bona-fide mCP stars. All in all, our final sample consisted of 41 spectroscopically confirmed mCP stars (39 of which were new discoveries), five candidate mCP stars and seven stars that do not exhibit peculiarities in the employed spectroscopic material.

Light curve stability was found to be a decisive criterion when searching for new mCP star candidates. While the light curves of the confirmed mCP stars remained stable throughout the observational time span, the seven non-CP objects displayed apparent changes in light curve shape (see Fig.\,\ref{develop}). An interesting result was that about one-quarter of the confirmed mCP stars exhibits a hitherto unobserved wealth of detail in their light curves indicative of complex surface structures. All new mCP stars are main sequence objects in various evolutionary stages in the mass range from 1.5\,\Ms\ to 4.0\,\Ms. For more details on our sample stars, we refer to \citet{Huemm18}.

\section{'Zoological garden' of \kepler\ mCP star light curves}

The accuracy of the employed photometry is typically 0.1\,mmag (ranging from 0.04 to 2.7\,mmag), but this cannot be achieved without further data treatment \citep{Huemm18}. Data from the \kepler\ pipeline are optimally accurate in the time interval of several days. In the scale of tens and hundreds of days, however, instrumental trends are encountered that considerably lower the theoretical accuracy limit (up to a factor of about 30). Although we were not able to entirely remove these instrumental effects, their impact can be considerably reduced by assuming that mCP star light curves are constant and devoid of sudden excursions. This assumption agrees with our current understanding of mCP stars and the finding that their light curves remain stable for decades.

\articlefigure[width=0.9\textwidth]{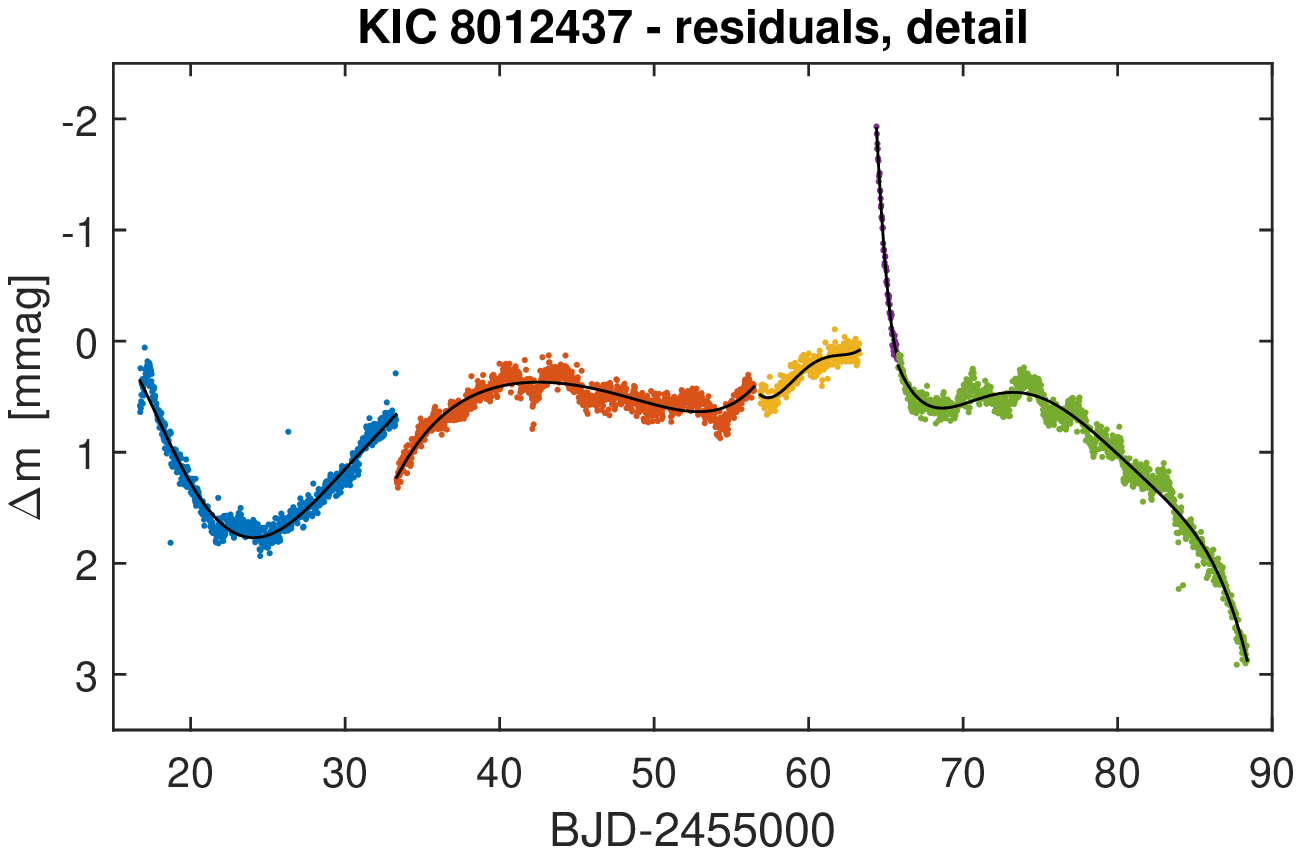}{resid}{Detailed view of the dependence of the individual residuals (coloured dots) on time in days, as exemplified by the case of KIC\,8012437. The full lines denote Chebyshev polynomials up to the sixth order that were fitted to the individual segments.}

By modeling the expected LC shape, we were able to separate intrinsic light variations from trends and artefacts. The instrument-related residuals were divided into several tens of unevenly long segments and fitted by Chebyshev polynomials of up to sixth order (Fig.\,\ref{resid}). After correcting the fits by robust regression, which eliminates outliers, we obtained a detrended LC with an accuracy about ten times higher. Several data segments still contained instrumental short-term oscillations that we were not able to remove, which precluded achieving the theoretical accuracy limit. Nevertheless, the results of our detrending procedure yielded mCP star light curves of unprecedented accuracy (Fig.\,\ref{normaly}).

\articlefigure[width=0.85\textwidth]{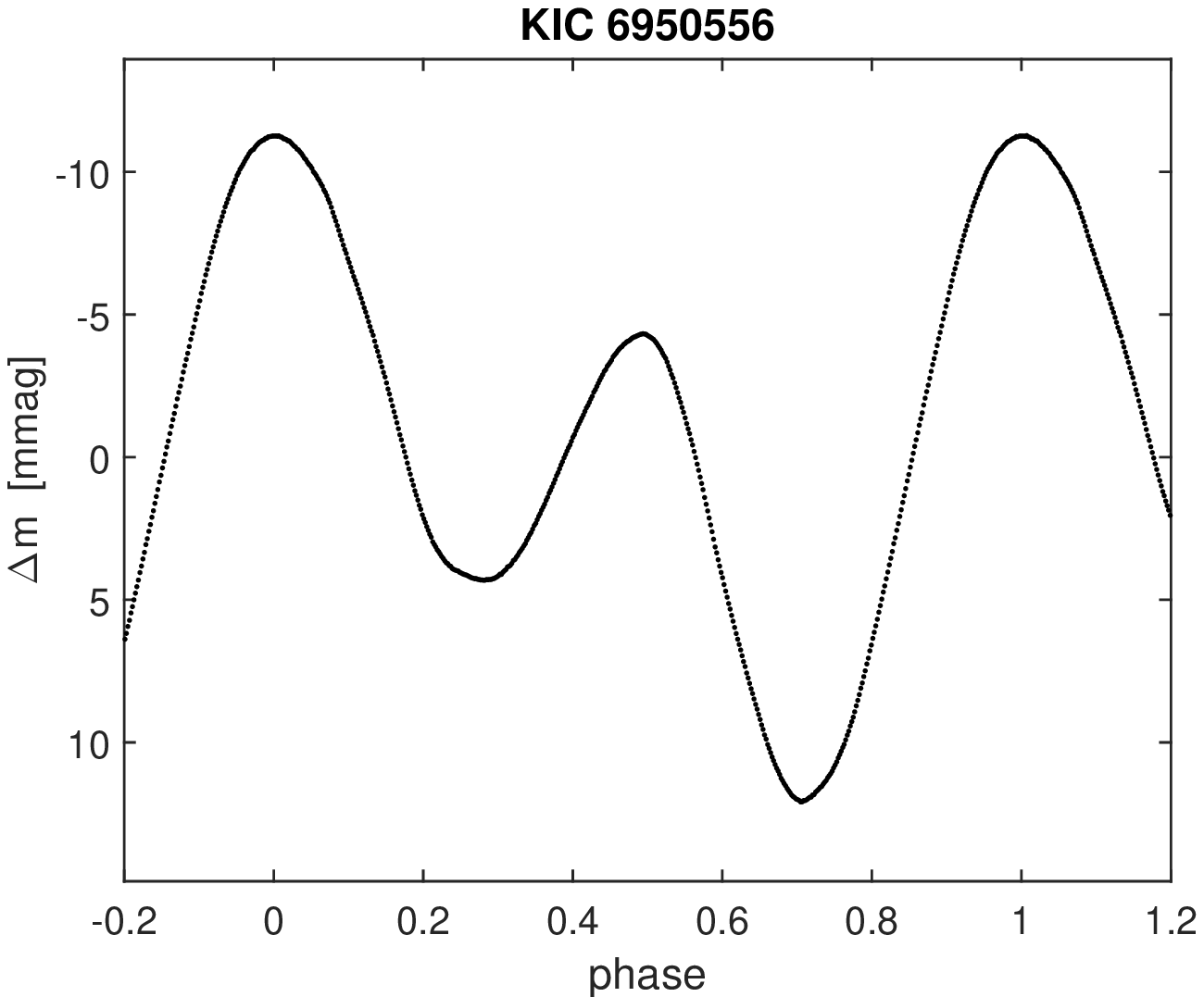}{normaly}{Light curve of the CP2 star KIC\,6950556 (A0V\,Si), phased with the rotation period of $P$\,=\,1.511\,785\,05(4)~d. Each point represents a median of 700 neighbouring detrended \kepler\ measurements.}

Each star displays a unique light curve shape. Single- and double-waved light curves clearly dominate (see Fig.\,\ref{kolac}), but some stars display a wealth of details (Fig.\,\ref{wealth}). This leads to the question of how to interpret the more complex light curves with the standard model of a rotating spotted star.

The detrended light curves can be described using harmonic polynomials of up to approximately the twentieth order. This indicates a complex surface brightness distribution, which nurtures suspicions that the roughly dipole-like magnetic field geometry in mCP stars does not determine the global spot appearance to such an extent as it was hitherto believed.

\articlefigure[width=0.75\textwidth]{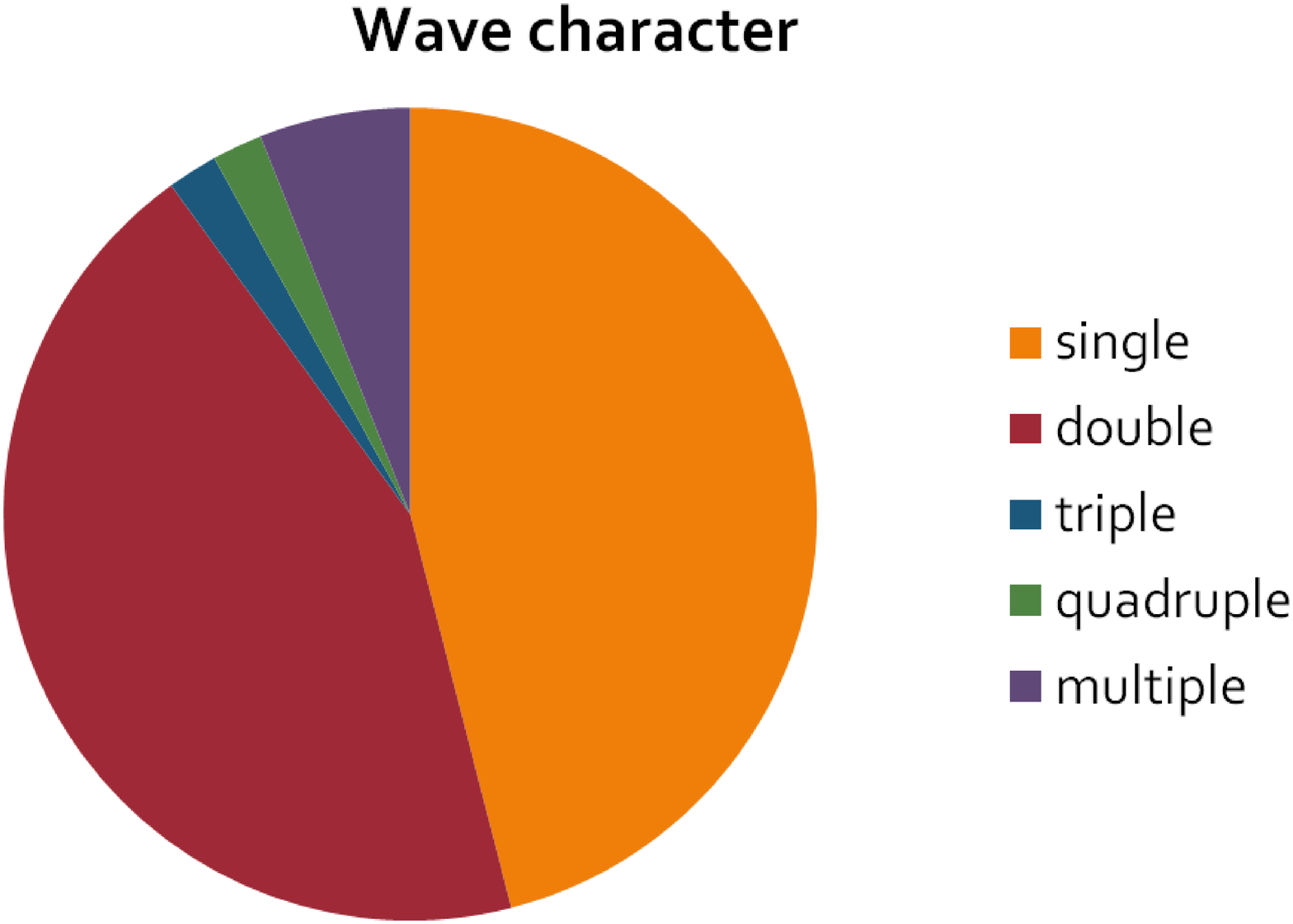}{kolac}{Breakdown on the shape of the light curves of the new mCP star sample. The most common shapes are the single wave ($N$\,=\,23) and the double wave ($N$\,=\,22). Nevertheless, the sample also contains several stars with more waves in their light curves ($N$\,=\,5).}

The median of the observed effective variability amplitudes is 17\,mmag. Amplitudes of a few milimagnitudes are quite common (see Fig.\,\ref{sloupce}, left panel), and it seems that there is no lower amplitude limit. The periods of the new mCP stars are known with excellent accuracy. With a median period of about three days, mCP stars as a group are generally slow rotators (see Fig.\,\ref{sloupce}, right panel). No correlation between effective amplitude and rotational period was observed.

\subsection{Candidate mCP stars / non-CP stars}

Most of the mCP star candidates identified with the criteria formulated in Sect.\,\ref{criteria} were confirmed by spectroscopy and exhibit various conspicuous peculiarities in their spectra. Nevertheless, some candidates remained without (new) spectroscopic observations (KIC\,2969628, 3326428, 6278403, 8362546, and 11560273) or did not show any peculiarities in the employed spectroscopic material \citep{Huemm18}.

The latter group consists of KIC\,5213466, 5727964, 8415109, 8569986, 10082844, 10550657, and 11671226. Interestingly, all of these stars display light curves compatible with rotational modulation but, in contrast to the mCP stars, also some kind of additional long-term variability \citep[][see also Fig.\,\ref{develop}, right panel]{Huemm18} that suggests classical starspots \citep{balona} rather than the longer-lived chemical spots. Time-resolved spectroscopy and photometry of these stars would be highly desirable to further investigate the origin of the observed light variations.

\articlefigure[width=0.75\textwidth]{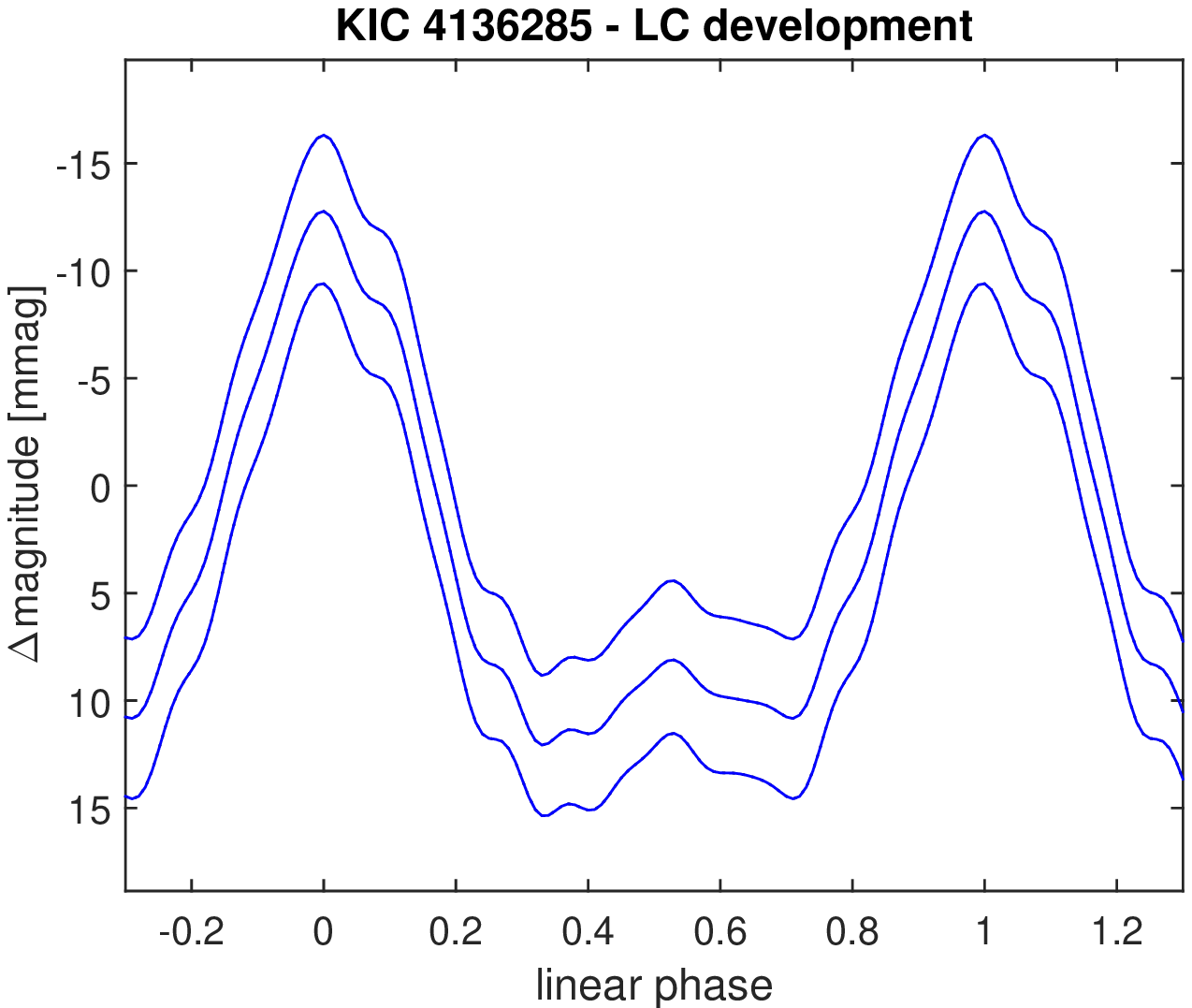}{wealth}{Light curve development of KIC\,4136285 \citep[spectral type B5, He weak;][]{RM09}. The time interval between subsequent mean light curves is 69 days. The star exhibits a highly complex but perfectly stable light curve throughout the covered time span.}

\articlefiguretwo{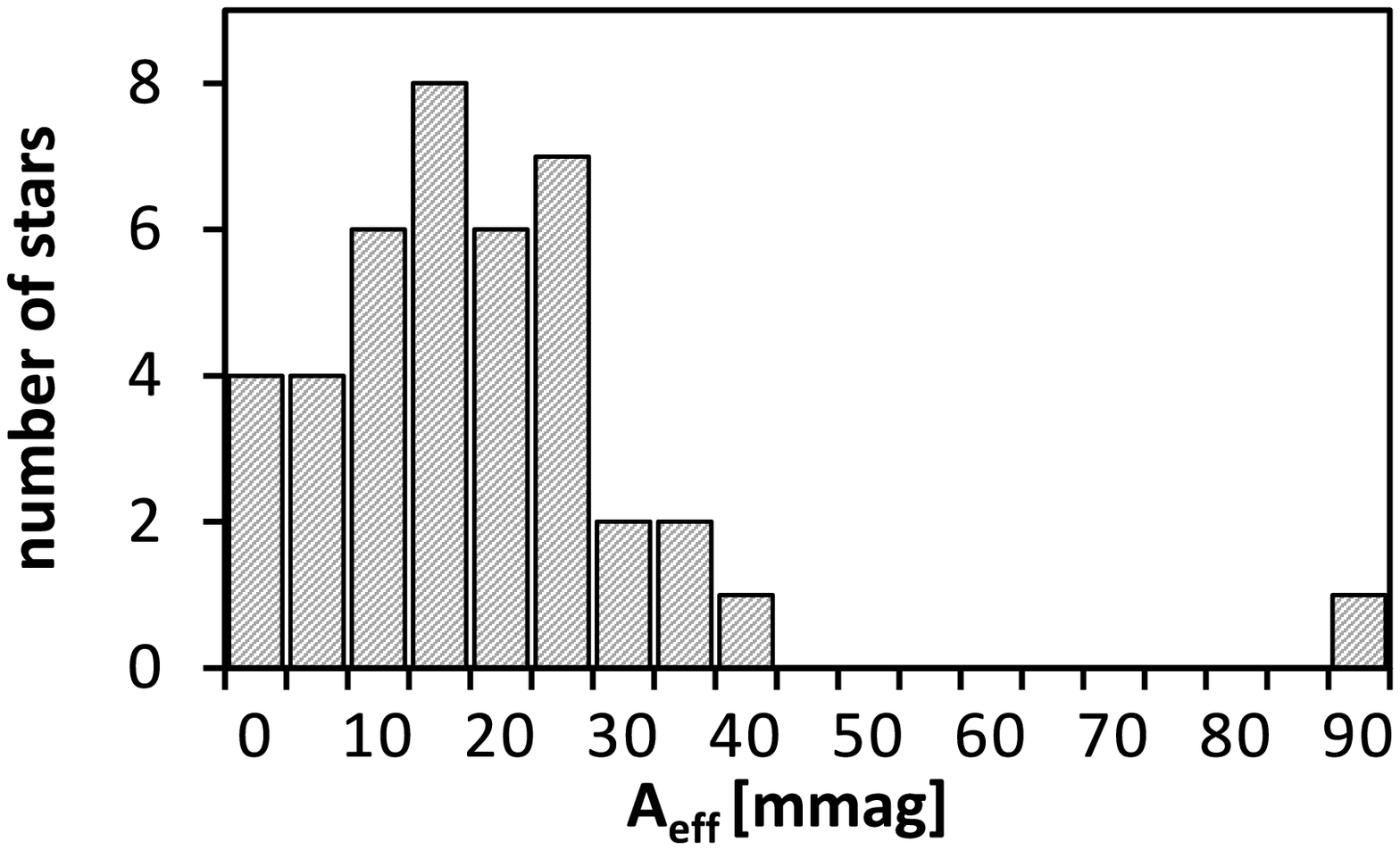}{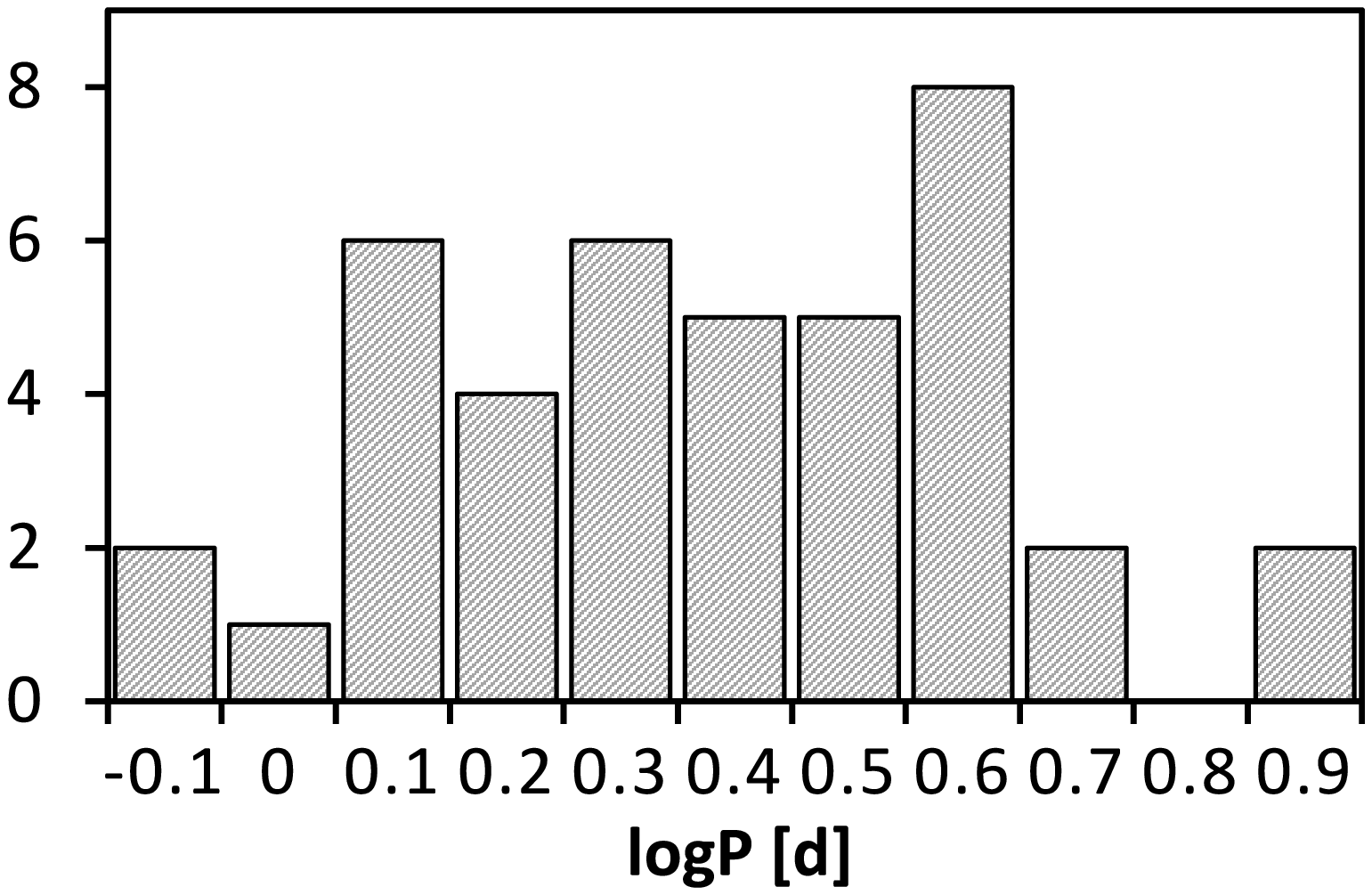}{sloupce}{Distribution of effective amplitudes (left panel) and rotational periods (right panel) among the 41 spectroscopically confirmed mCP stars.}

\section{Principal Component Analysis of the \kepler\ mCP star light curves}

For an adequate description of the \kepler\ mCP star light curves, harmonic polynomials of rather high orders are needed. However, the rather smooth aspect of the light curves renders them targets for Principal Component Analysis (PCA). Applying this method, we reached the conclusion that all studied \kepler\ mCP star light curves (including the details, see Fig.\,\ref{PCALC}) can be satisfactorily fitted by the linear combination of only eleven basic light curves. This is due to the fact that the individual terms of the harmonic decomposition are mutually bound, which means that their description using harmonic polynomials is internally redundant. We can therefore conclude that even very precise photometry does not allow to distinguish more than eleven detailed components -- photometric spots or their conglomerations -- on the surface of a rotating spotted star.

\articlefiguretwo{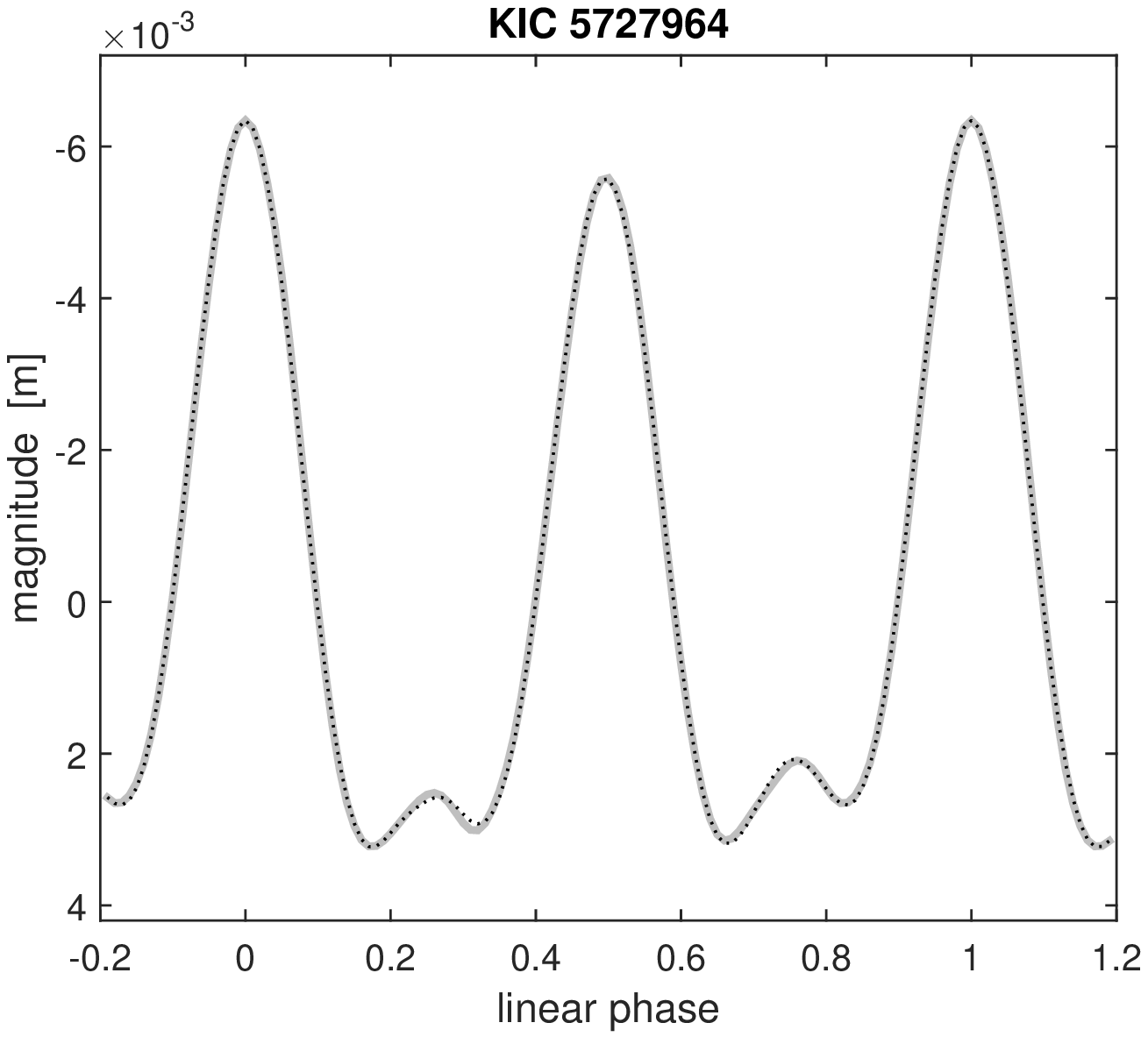}{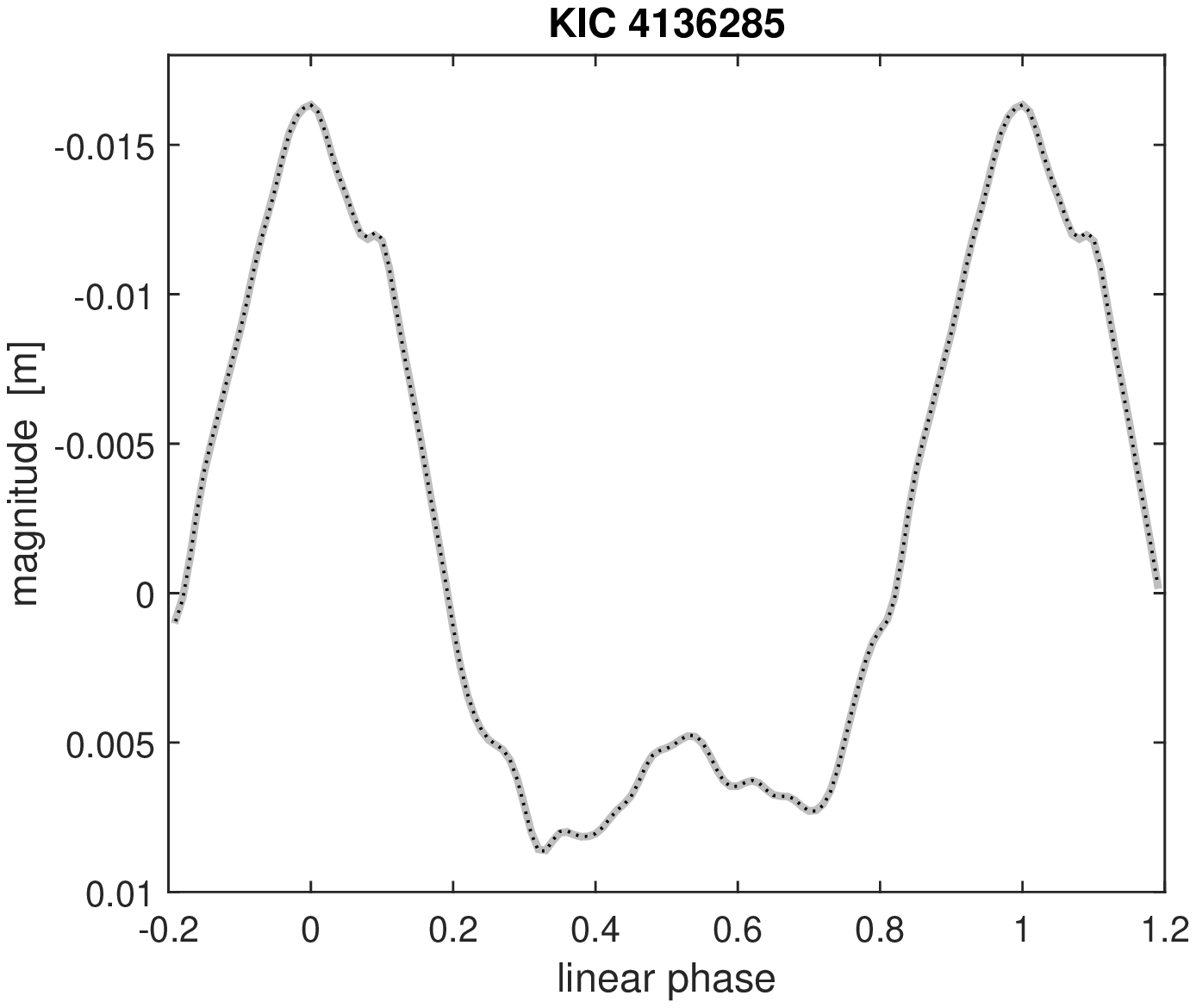}{PCALC}{The eleven component PCA fits of the complex light curves of KIC\,5727964 (left) and\,KIC 4136285 (right). The grey line corresponds to the mean light curve, black dots are the PCA fit.}

\section{Outlook}

Although the stability of the light curves of the new confirmed mCP stars is remarkable, careful inspection of the non-detrended light curves proves the presence of very weak secular variations in the shape of some of them. Future research will be based on the detrended light curves and may reveal more details. The same holds true for phase changes, which were detected in at least one fifth of the new mCP stars. These variations are probably caused by the presence of companions of various masses (from brown dwarfs to hot MS stars) with orbital periods comparable to the period of the \kepler\ observations. This fact may have serious consequences regarding the explanation why mCP stars are rotating so slowly.

\acknowledgements This research was supported by grant GA\,\v{C}R 16-01116S.

\bibliography{Overview}

\begin{thebibliography}{}
\expandafter\ifx\csname natexlab\endcsname\relax\def\natexlab#1{#1}\fi
\expandafter\ifx\csname url\endcsname\relax
  \def\url#1{\texttt{#1}}\fi
\expandafter\ifx\csname urlprefix\endcsname\relax\def\urlprefix{URL }\fi
\providecommand{\eprint}[2][]{eprint: #2}

\bibitem[{{Balona}(2017)}]{balona}
{Balona}, L.~A. 2017, \mnras, 467, 1830

\bibitem[{{Bychkov} et~al.(2009){Bychkov}, {Bychkova}, \& {Madej}}]{Bych09}
{Bychkov}, V.~D., {Bychkova}, L.~V., \& {Madej}, J. 2009, \mnras, 394, 1338

\bibitem[{{Deutsch}(1970)}]{Deutsch70}
{Deutsch}, A.~J. 1970, \apj, 159, 985

\bibitem[{{Dukes} \& {Adelman}(2018)}]{Dukes18}
{Dukes}, R.~J., Jr., \& {Adelman}, S.~J. 2018, \pasp, 130, 044202

\bibitem[{{H{\"u}mmerich} et~al.(2018){H{\"u}mmerich}, {Mikul{\'a}{\v s}ek},
  {Paunzen}, {Bernhard}, {Jan{\'{\i}}k}, {Yakunin}, {Pribulla}, {Va{\v n}ko},
  \& {Mat{\v e}chov{\'a}}}]{Huemm18}
{H{\"u}mmerich}, S., {Mikul{\'a}{\v s}ek}, Z., {Paunzen}, E., {Bernhard}, K.,
  {Jan{\'{\i}}k}, J., {Yakunin}, I.~A., {Pribulla}, T., {Va{\v n}ko}, M., \&
  {Mat{\v e}chov{\'a}}, L. 2018, \aap, 619, A98. \eprint{1808.05669}

\bibitem[{{Krti{\v c}ka} et~al.(2012){Krti{\v c}ka}, {Mikul{\'a}{\v s}ek},
  {L{\"u}ftinger}, {Shulyak}, {Zverko}, {{\v Z}i{\v z}{\v n}ovsk{\'y}}, \&
  {Sokolov}}]{Krti12}
{Krti{\v c}ka}, J., {Mikul{\'a}{\v s}ek}, Z., {L{\"u}ftinger}, T., {Shulyak},
  D., {Zverko}, J., {{\v Z}i{\v z}{\v n}ovsk{\'y}}, J., \& {Sokolov}, N.~A.
  2012, \aap, 537, A14. \eprint{1111.2746}

\bibitem[{{Maitzen}(1984)}]{maitzen}
{Maitzen}, H.~M. 1984, \aap, 138, 493

\bibitem[{{Mathys} \& {Manfroid}(1985)}]{mathys}
{Mathys}, G., \& {Manfroid}, J. 1985, \aaps, 60, 17

\bibitem[{{Mikul\'a\v{s}ek} et~al.(2007){Mikul\'a\v{s}ek}, {Zverko}, {Krticka},
  {Janik}, {Ziznovsky}, \& {Zejda}}]{mikzoo}
{Mikul\'a\v{s}ek}, Z., {Zverko}, J., {Krticka}, J., {Janik}, J., {Ziznovsky},
  J., \& {Zejda}, M. 2007, arXiv Astrophysics e-prints.
  \eprint{astro-ph/0703521}

\bibitem[{{Paunzen} et~al.(2005){Paunzen}, {St{\"u}tz}, \& {Maitzen}}]{Paunz05}
{Paunzen}, E., {St{\"u}tz}, C., \& {Maitzen}, H.~M. 2005, \aap, 441, 631.
  \eprint{astro-ph/0507124}

\bibitem[{{Preston}(1974)}]{preston}
{Preston}, G.~W. 1974, \araa, 12, 257

\bibitem[{{Renson} \& {Manfroid}(2009)}]{RM09}
{Renson}, P., \& {Manfroid}, J. 2009, \aap, 498, 961

\bibitem[{{{\v Z}i{\v z}\v{ n}ovsk{\'y}}(1994)}]{Zizn94}
{{\v Z}i{\v z}\v{ n}ovsk{\'y}}, J. 1994, in Chemically Peculiar and Magnetic
  Stars, edited by J.~{Zverko}, \& J.~{Ziznovsky}, 155

\end{thebibliography}

\end{document}